\documentclass{article}
\usepackage{cite}
\usepackage{url}
\usepackage{times}
\usepackage{siunitx}
\usepackage{alltt}
\usepackage{graphicx}
\usepackage{listings}

\DeclareFixedFont{\ttb}{T1}{txtt}{bx}{n}{8} 
\DeclareFixedFont{\ttm}{T1}{txtt}{m}{n}{8}  

\usepackage{color}
\definecolor{bg}{rgb}{0.96,0.96,0.85}
\definecolor{deepblue}{rgb}{0,0,0.5}
\definecolor{deepred}{rgb}{0.6,0,0}
\definecolor{deepgreen}{rgb}{0,0.5,0}

\usepackage{xcolor}

\definecolor{gray}{gray}{0.5}
\colorlet{commentcolour}{green!50!black}

\colorlet{stringcolour}{red!60!black}
\colorlet{keywordcolour}{magenta!90!black}
\colorlet{exceptioncolour}{yellow!50!red}
\colorlet{commandcolour}{blue!60!black}
\colorlet{numpycolour}{blue!60!green}
\colorlet{literatecolour}{magenta!90!black}
\colorlet{promptcolour}{green!50!black}
\colorlet{specmethodcolour}{violet}
\colorlet{indendifiercolour}{green!70!white}

\newcommand{\literatecolour}{\textcolor{literatecolour}}

\newcommand\pythonstyle{\lstset{
language=python,
showtabs=true,
tab=,
tabsize=2,
basicstyle=\ttfamily\footnotesize,
stringstyle=\color{stringcolour},
showstringspaces=false,
alsoletter={1234567890},
otherkeywords={\ , \}, \{, \%, \&, \|},
keywordstyle=\color{keywordcolour}\bfseries,
emph={and,break,class,continue,def,yield,del,elif ,else,%
except,exec,finally,for,from,global,if,import,in,%
lambda,not,or,pass,print,raise,return,try,while,assert},
emphstyle=\color{blue}\bfseries,
emph={[2]True, False, None},
emphstyle=[2]\color{keywordcolour},
emph={[3]object,type,isinstance,copy,deepcopy,zip,enumerate,reversed,list,len,dict,tuple,xrange,append,execfile,real,imag,reduce,str,repr},
emphstyle=[3]\color{commandcolour},
emph={Exception,NameError,IndexError,SyntaxError,TypeError,ValueError,OverflowError,ZeroDivisionError},
emphstyle=\color{exceptioncolour}\bfseries,
morestring=[s]{"""}{"""},
morestring=[s]{'''}{'''},
commentstyle=\color{commentcolour}\slshape,
emph={[4]ode, fsolve, sqrt, exp, sin, cos, arccos, pi,  array, norm, solve, dot, arange, , isscalar, max, sum, flatten, shape, reshape, find, any, all, abs, linspace, legend, quad, polyval,polyfit, hstack, concatenate,vstack,column_stack,empty,zeros,ones,rand,vander,grid,pcolor,eig,eigs,eigvals,svd,qr,tan,det,logspace,roll,min,mean,cumsum,cumprod,diff,vectorize,lstsq,cla,eye,xlabel,ylabel,squeeze},
emphstyle=[4]\color{numpycolour},
emph={[5]__init__,__add__,__mul__,__div__,__sub__,__call__,__getitem__,__setitem__,__eq__,__ne__,__nonzero__,__rmul__,__radd__,__repr__,__str__,__get__,__truediv__,__pow__,__name__,__future__,__all__},
emphstyle=[5]\color{specmethodcolour},
emph={[6]assert,range,yield},
emphstyle=[6]\color{keywordcolour}\bfseries,
literate=*%
{:}{{\literatecolour:}}{1}%
{=}{{\literatecolour=}}{1}%
{-}{{\literatecolour-}}{1}%
{+}{{\literatecolour+}}{1}%
{*}{{\literatecolour*}}{1}%
{/}{{\literatecolour/}}{1}%
{!}{{\literatecolour!}}{1}%
{[}{{\literatecolour[}}{1}%
{]}{{\literatecolour]}}{1}%
{<}{{\literatecolour<}}{1}%
{>}{{\literatecolour>}}{1}%
{>>>}{{\textcolor{promptcolour}{>>>}}}{1}%
,%
breaklines=true,
breakatwhitespace= true,
aboveskip=1ex,
frame=trbl,
rulecolor=\color{black!40},
backgroundcolor=\color{yellow!10}
}}

\lstnewenvironment{python}[1][]
{
  \pythonstyle
  \lstset{#1}
}
{}

\newcommand\pythoninline[1]{{\pythonstyle\lstinline!#1!}}

\usepackage{amsmath, amssymb, amscd}
\usepackage{algorithm, algorithmic} 
\usepackage[top=1in, bottom=1in, left=1in, right=1in]{geometry}
\newcommand{\sups}[1]{\ensuremath{^{\textrm{#1}}}}

\makeatletter
\renewcommand\@biblabel[1]{\textbullet}
\makeatother

\begin{document}


\title{Development of a finite element firn densification model for converting volume changes to mass changes}
\author{Evan Cummings \and Jesse Johnson \and Douglas Brinkerhoff}
\maketitle
\begin{center}
\includegraphics[width=4.455666122085252in]{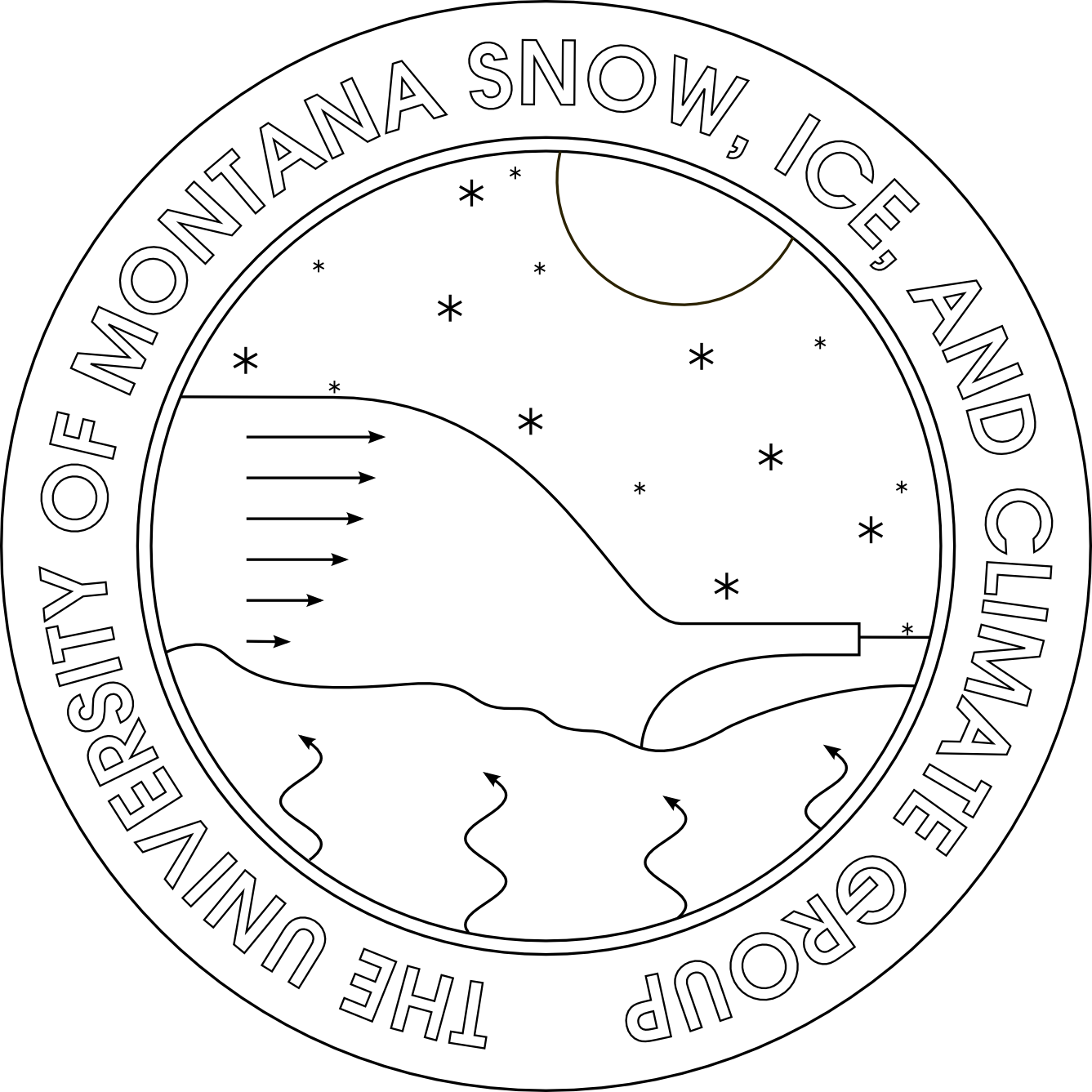}
\end{center}

\twocolumn

\section{Introduction}

In regions where ice sheets are increasing in mass, there is a 50-200 m layer of old snow called \emph{firn} which does not melt in the summer months.  The density of firn tracks the transformation of snow into glacial ice at approximately 917 kg m\sups{-3}.  The process of firn densification is important in at least two ways: 1) it can be a dominant component in the observed rate of change of the surface elevation, and 2) storage of liquid water in the lower density firn layer is now considered a critical component in the mass balance of ice sheets.  If the rate of change of surface elevation can be equated with the rate of change in the mass of the ice sheet, we would have an excellent means of monitoring ice sheet mass balance.  However, knowledge of firn densification rates is needed to make the inference of mass rate of change from volume rate of change.

Several firn models have been created for areas without melt.  \emph{Herron and Langway} [1980] developed a firn densification model based on Arrhenius-type equations with variable rate constants, and found that the densification rate increased suddenly around 550 kg m\sups{-3}.  \emph{Zwally and Li} [2002] expanded upon this model and found an alternate temperature-dependent value for the rate constant.  \emph{Arthern et al.} [2010] developed yet another set of equations based on their in situ measurements of Antarctic snow compaction, while \emph{Ligtenberg et al.} [2011] modified the Arthern parametrization to better fit areas with a higher average annual temperature. 

We have reformulated these models with the finite-element software package FEniCS and integrated them with an enthalpy-formulation proposed by \emph{Aschwanden et al.} [2012].  This integration allows us to account for the melting and subsequent re-freezing of firn layers into ice \emph{lenses}.

\section{Temperature Solution}

We begin with the standard heat-transport equation as explained by \emph{Patterson} [2001]
\begin{align}
  \rho c_i \frac{\partial T}{\partial t} = 
    k_i \frac{\partial^2 T}{\partial z^2} +
    \left( \frac{dk_i}{dt} - \rho c_i w \right) \frac{\partial T}{\partial z}
\end{align}
with heat sources from the deformation of ice omitted, $\rho$ density, $c_i$ heat capacity, $k_i$ thermal conductivity, $w$ vertical velocity, and $T$ temperature of firn.  To solve the total derivative $dk_i/dt$ we must apply the chain rule
\begin{align*}
  \frac{dk_i}{dz} = 
  \frac{\partial k_i}{\partial \rho} \frac{\partial \rho}{\partial z} + 
  \frac{\partial k_i}{\partial T} \frac{\partial T}{\partial z}.
\end{align*}
The thermal conductivity of firn is defined by \emph{Arthern et al.}, [1998] as
\begin{align}
  k_i = 2.1 \left(\frac{\rho}{\rho_i}\right)^2,
\end{align}
from which we find
\begin{align*}
  \frac{\partial k_i}{\partial \rho} = 
    4.2 \frac{\rho}{\rho_i^2}
\end{align*}
and
\begin{align*}
  \frac{\partial k_i}{\partial T} = 
    \frac{4.2}{\rho_i^2} \left( \frac{\partial \rho}{\partial T} \right).
\end{align*}
\emph{Patterson} [2001] defined the heat capacity $c_i$ of firn with the equation
\begin{align}
  c_i = 152.5 + 7.122 T.
\end{align}
However, we use a constant heat capacity for simplification.  Contained within the same literature is an expression $\rho$ in terms of $T$ which leads to the equation:
\begin{align}
  \frac{\partial \rho}{\partial T} = 
    (\SI{5.6e-2}) \exp ((\SI{-5.7e-3})T).
\end{align}

\section{Velocity Solution}

The vertical velocity $w$ can be found by integrating the densification rate:
\begin{align*}
  w(z,t) &= \int \frac{1}{\rho(z)} \frac{d\rho(z)}{dt}\ dz.
\end{align*}
Taking the derivative with respect to $z$,
\begin{align}
  \frac{\partial}{\partial z} w &= \frac{\partial}{\partial z} \int \frac{1}{\rho} \frac{d\rho}{dt}\ dz \notag \\
  \frac{\partial w}{\partial z} &= \frac{1}{\rho} \frac{d\rho}{dt}.
\end{align}

\section{Density Solution}

The densification process is defined with the material derivative
\begin{align}
  \frac{d \rho}{dt} = \frac{\partial \rho}{\partial t} + 
  w\frac{\partial \rho}{\partial z}.
\end{align}
\emph{Arthern et al.} [2010] described this derivative differently for density values above and below a critical value, $\rho_m$:  
\begin{align}
  \frac{d \rho}{dt} = 
  \begin{cases}
   c_0(\rho_i - \rho), &\rho \leq \rho_m\\
   c_1(\rho_i - \rho), &\rho > \rho_m
  \end{cases}.
\end{align}
\emph{Zwally and Li} [2002] defined a single multiplying constant $c$ with an Arrhenius-type relation
\begin{align*}
  c_0 = c_1 = 
  \dot{b} \beta(T)\left(\frac{\rho_i}{\rho_w}\right)
  K_{0G}(T)\exp \left( -\frac{E(T)}{RT} \right),
\end{align*}
with $K_{0G}(T) \exp(-E(T)/(RT)) = 8.36T^{-2.061}$ as described in \emph{Reeh} [2008], average annual accumulation rate $\dot{b}$ in units of kg m\sups{-2} s\sups{-1}, and $\beta(T)$ a unitless smoothing function to match a desired density rate.  \emph{Arthern et al} [2010] developed a semi-empirical formula by coupling the rate equations for Nabarro-Herring creep and normal grain-growth: 
\begin{align}
  \begin{cases}
    c_0 = M_0 \dot{b}g\frac{k_{c0}}{k_g}\exp\left(-\frac{E_c}{RT} + 
          \frac{E_g}{RT_{avg}}\right)\\
    c_1 = M_1 \dot{b}g\frac{k_{c1}}{k_g}\exp\left(-\frac{E_c}{RT} + 
          \frac{E_g}{RT_{avg}}\right)
  \end{cases},
\end{align}
with the creep coefficients defined as
\begin{align*}
  \begin{cases}
    k_{c0} = \text{\SI{9.2e-9} m\sups{3} s kg\sups{-1}} \\
    k_{c1} = \text{\SI{3.7e-9} m\sups{3} s kg\sups{-1}}  
  \end{cases}
\end{align*}
and $M$ defined in \emph{Ligtenberg et al.} [2011] to better fit with observed densification rates in higher-temperature environments:
\begin{align*}
  \begin{cases}
    M_0 = 2.366 - 0.293\ln(\dot{b}*\SI{1e3})\\
    M_1 = 1.435 - 0.151\ln(\dot{b}*\SI{1e3})
  \end{cases}.
\end{align*}

\section{Enthalpy Solution}

As stated in \emph{Aschwanden et al.} [2012], we take 'enthalpy' to be synonymous with 'internal energy' due to the exclusion of work done with changing volume.  The equation used here is the shallow-enthalpy:
\begin{align}
  \rho \frac{\partial H}{\partial t} = \frac{\partial}{\partial z} 
    \left( 
      \begin{Bmatrix}
        K_i, &\text{Cold}\\
        K_0, &\text{Temperate}
      \end{Bmatrix}
      \frac{\partial H}{\partial z} 
    \right) + w \rho \frac{\partial H}{\partial z}.
\end{align}
Strain heating has been neglected and the advective term $w \rho\ \partial H / \partial z$ has been added.  The coefficient for cold ice is 
\begin{align*}
  K_i = \frac{k_i}{c_i},
\end{align*}
and the coefficient for temperate ice is
\begin{align*}
  K_0 = \frac{1}{10}K_i.
\end{align*}
Temperate firn is defined as firn with $H > H_s$, cold firn with $H \leq H_s$, where
\begin{align*}
  H_s = \int_{T_0}^{T_w}{c_i(T)}dT,
\end{align*}
with $T_w = 273.15$ K and $T_0 = 0.0$ K.  The enthalpy can be found with a constant heat capacity of $2009$ J kg\sups{-1} K\sups{-1} by the linear equation
\begin{align}
  H = 
  \begin{cases}
    c_i(T - T_0), &\text{where } T \leq T_w\\
    c_i(T_w - T_0) + \omega L_f,  &\text{where } T > T_w
  \end{cases}
\end{align}
where $L_f$ is the latent heat of fusion and $\omega$ represents the water content percentage of firn given by
\begin{align}
  \omega L_f = H - c_i(T_w - T_0).
\end{align}
Temperature may be derived from enthalpy easily:
\begin{align}
  T = \frac{H}{c_i}.
\end{align}
The density of the firn column changes with the percentage of water content:
\begin{align*}
  \rho^n = 
  \begin{cases}
    \rho^{n-1} + \Delta\omega(\rho_w - \rho_i)\ \text{kg m\sups{-3}},  
      &\Delta\omega \leq 0\\
    \rho^{n-1} + \Delta\omega\rho_w\ \text{kg m\sups{-3}}, 
      &\Delta\omega > 0
  \end{cases},
\end{align*}
where $\Delta\omega = \omega^n - \omega^{n-1}$ is the change in water content and superscripts refer to the time index.  This has the effect of adding water to the firn column and refreezing the portion of firn with decreasing water content.  The surface-density at time index $n+1$ can be described as: 
\begin{align*}
  \rho_s^{n+1} = \rho_{\dot{b}}^{n+1} d_p + \rho_s^{n} (1 - d_p),
\end{align*}
where
\begin{align*}
  \rho_{\dot{b}}^{n+1} &= \rho_{\dot{b}}^{n-1} + \Delta \omega_s \rho_{\dot{b}}^n,\\
  \Delta \omega_s &= \omega_s^{n} - \omega_s^{n-1},\\
  d_p &= \frac{d_n}{l_s},\\
  d_n &= w_s\Delta t,\text{ and}\\
  l_s &\text{ is the length of the surface node.}
\end{align*}
If $T_s \geq T_w$, the density of surface snow while taking into account re-freezing is simulated making
\begin{align*}
  \rho_{\dot{b}}^n = 
  \begin{cases}
    \rho_w - \rho_i\ \text{kg m\sups{-3}},  &\Delta\omega_s < 0\\
    \rho_w\ \text{kg m\sups{-3}}, &\Delta\omega_s > 0
  \end{cases},
\end{align*}
but when $T_s < T_w$, $\rho_{\dot{b}}^n$ is simply made to be $\rho_s$.
\begin{figure}[H]
	\centering
		\includegraphics[width=0.42\textwidth]{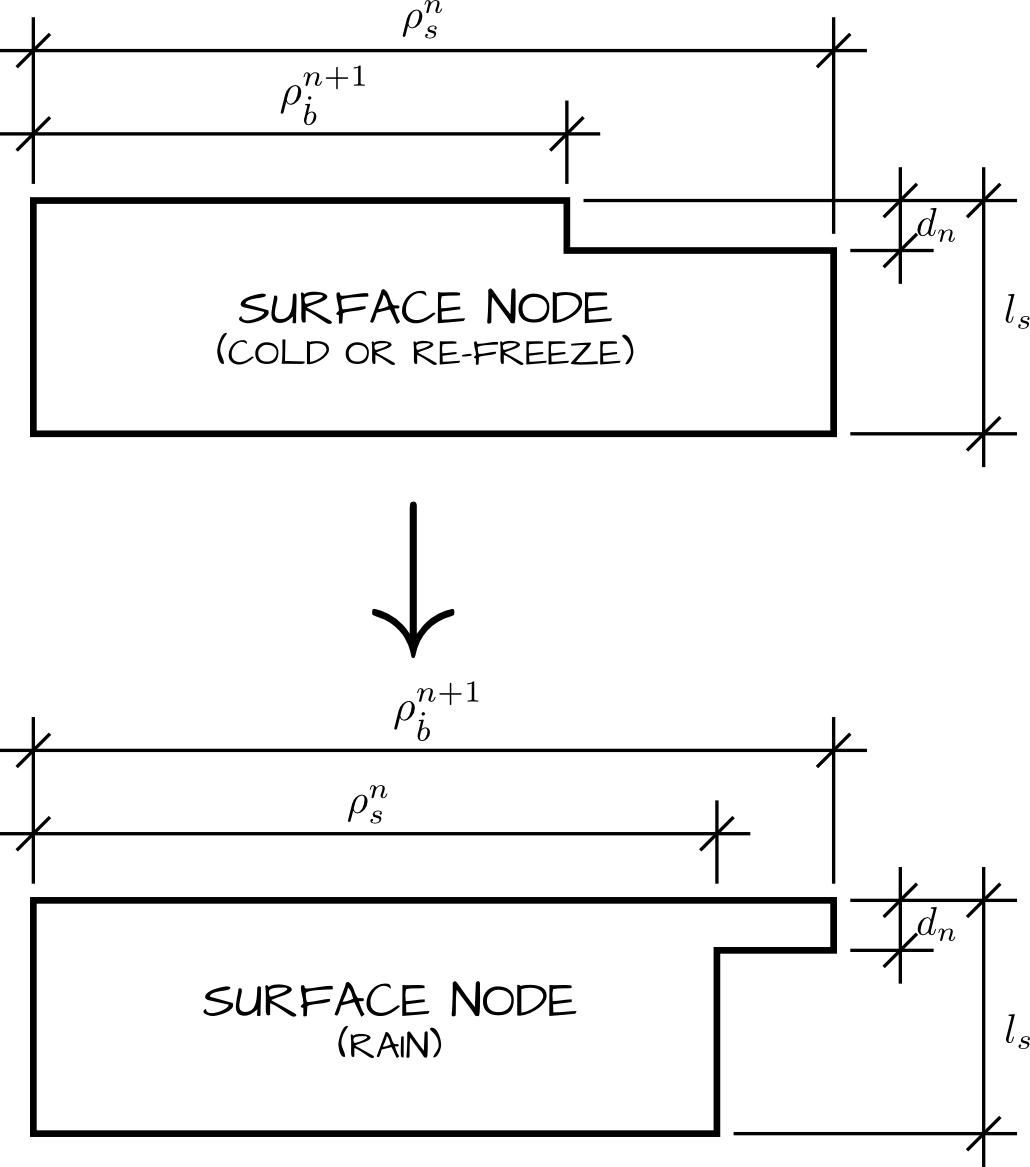}
	\label{fig:500 year orbit}
	\caption{Evolution of surface node density}
\end{figure}

\section{Finite Element Method}

This section will focus on the enthalpy solution.  Solving these equations with FEniCS is done with Galerkin's method and requires finding the weak formulation of equation (9):
\begin{align*}
    f_H = 0 =
    &\int_{\Omega} 
      \begin{Bmatrix}
        K_i\\
        K_0
      \end{Bmatrix}
      \nabla^2 H \psi\ d \Omega 
    + \int_{\Omega}w \rho \nabla H \psi\ d \Omega \\
    &- \int_{\Omega} {\rho \frac{\partial H}{\partial t}} \psi\ d \Omega.
\end{align*}
Here we have integrated the enthalpy equation over the entire domain, $\Omega$, and multiplied by a test function $\psi$.  Note that we have used the gradient operator.  Although our simulation is one-dimensional, it can easily be extended to two or three and thus we use the gradient operator for clarity of notation.  For the purposes of this paper, take $\nabla{\alpha} = \frac{\partial \alpha}{\partial z}$.  After integrating the diffusive term by parts the above equation becomes
\begin{align*}
    f_H =
    & \int_{\Gamma} 
        \begin{Bmatrix}
          K_i\\
          K_0
        \end{Bmatrix}
        \nabla H \psi\ d \Gamma
    - \int_{\Omega} 
        \begin{Bmatrix}
          K_i\\
          K_0
        \end{Bmatrix}
        \nabla H \nabla \psi\ d \Omega \\ 
    &+ \int_{\Omega}w \rho \nabla H \psi\ d \Omega
    - \int_{\Omega} {\rho \frac{\partial H}{\partial t}} \psi\ d \Omega.
\end{align*}
The boundary surfaces ($\Gamma$) of our problem are all Dirichlet or free-Neumann, and thus the boundary term is required to vanish, i.e., $\int_{\Gamma} \nabla H \psi\ d \Gamma = 0$.  Our final residual is
\begin{align}
    f_H =
    &\int_{\Omega}w \rho \nabla H \psi\ d \Omega
    - \int_{\Omega} 
        \begin{Bmatrix}
          K_i\\
          K_0
        \end{Bmatrix}
        \nabla H \nabla \psi\ d \Omega \notag \\ 
    &- \int_{\Omega} {\rho \frac{\partial H}{\partial t}} \psi\ d \Omega.
\end{align}

We can discretize the enthalpy time-differential with the second-order accurate formula
\begin{align*}
  \frac{\partial H}{\partial t} = \frac{H - H^{k-1}}{\Delta t} = \theta H^{k}  + (1-\theta) H^{k-1}
\end{align*}
with superscripts referring to time index.  Using a $\theta$-scheme as given by \emph{Logg et al.} [2011] and \emph{Zienkiewicz \& Taylor} [2005], $H_{mid} = \theta H^{k}  + (1-\theta) H^{k-1}$, where $\theta \in [0,1]$ is a weighting factor chosen from: 
\begin{align*}
    \theta = 
    \begin{cases}
      1,       & \text{Backwards-Euler}\\
      0.667,   & \text{Galerkin}\\
      0.878,   & \text{Liniger}\\
      0.5,     & \text{Crank-Nicolson}\\
      0,       & \text{Forward-Euler}
    \end{cases}
\end{align*}
The entire enthalpy residual can be represented in FEniCS as:\par
\begin{python}
theta = 0.5
H_mid = theta*H + (1 - theta)*H_1
f_H   = rho*w*H_mid.dx(0)*psi*dx - \
        k/c*Kcoef * \ 
        inner(H_mid.dx(0), psi.dx(0))*dx - \
        rho*(H - H_1)/dt*psi*dx
\end{python}
The variable \pythoninline{Kcoef} is a coefficient vector which will be updated dynamically depending on the temperature of firn (either $1.0$ or $0.1$ corresponding to $K_i$ and $K_0$).

The weak form for density is found similarly, with a upwinding necessary to eliminate artifacts due to the sudden increase in density where ice lenses formed.  The method used here is the Streamline-Upwind-Petrov-Galerkin (SUPG) method:
\begin{align*}
    \hat{\phi} = \phi + \frac{h}{2||w||} w \cdot \nabla{\phi},
\end{align*}
where $h$ is the cellsize.  The density residual of equation (6) after integration by parts becomes:
\begin{align}
  f_{\rho} = 
    \int_{\Omega} \frac{\partial \rho}{\partial t}\phi\ d \Omega + 
    \int_{\Omega} w\nabla \rho \hat{\phi}\ d \Omega -
    \int_{\Omega}\frac{d \rho}{dt}\hat{\phi}\ d \Omega.
\end{align}
With the partial-time differential of density defined identically to the enthalpy equation, this can be represented in FEniCS including the \emph{Arthern et al.} [2010] densification equation as:\par
\begin{python}
vnorm   = sqrt(dot(w, w) + 1e-10)
cellh   = CellSize(mesh)
phihat  = phi+cellh/(2*vnorm)*dot(w,grad(phi))
c       = b*g*rhoCoef/kg * 
          exp(-Ec/(R*T) + Eg/(R*Tavg))
drhodt  = c*(rhoi - rho)
theta   = 0.878
rho_mid = theta*rho + (1 - theta)*rho_1
f_rho   = (rho - rho_1)/dt*phi*dx - 
          (drhodt - w*grad(rho_mid))*phihat*dx
\end{python}
The variable \pythoninline{rhoCoef} is another dynamically updated coefficient vector and is either $k_{c0}$ or $k_{c1}$ depending upon the density at the node.  We have chosen $\theta = 0.878$ corresponding to the semi-implicit first order time stepping Liniger method; this is needed due to the jump condition at $\rho_m$.

In order to find the vertical velocity, equation (5) must be solved.  The weak formulation of the residual is:
\begin{align}
  f_{w} = 
    \int_{\Omega} \rho \nabla{w} \eta\ d\Omega + \int_{\Omega} \frac{d\rho}{dt} \eta\ d\Omega,
\end{align}
and is created with FEniCS by:
\begin{python}
theta = 0.878
w_mid = theta*w + (1 - theta)*w_1
f_w   = rho*grad(w_mid)*eta*dx + drhodt*eta*dx
\end{python}
We chose $\theta=0.878$ again due to the jump discontinuity at $\rho_m$.

We can define the function space for the entire non-linear problem as 
\begin{align*}
    U = \Omega \times \Omega \times \Omega,
\end{align*}
with corresponding trial and test functions respectively defined as
\begin{align*}
    d_h, j \subset U.
\end{align*}
The test functions for each function can now be described as
\begin{align*}
    \psi, \phi, \eta \subset j.
\end{align*}
In FEniCS these spaces can be defined by this:
\begin{python}
mesh        = IntervalMesh(n, zb, zs)
V           = FunctionSpace(mesh,'Lagrange',1)
MV          = MixedFunctionSpace([V, V, V])
h           = Function(MV)
H,rho,w     = split(h)    
dh          = TrialFunction(MV)
dH,drho,dw  = split(dh)
j           = TestFunction(MV)
psi,phi,eta = split(j)
\end{python}
The variable \pythoninline{zb} is the z-position of the base of the firn column which does not change and \pythoninline{n} is the number of nodes; the \pythoninline{mesh} variable defines the spacial dimensions of the system to be solved and is here created in one dimension.  The mesh may also be created in three dimensions if desired, or made to fit a custom grid.  The variables \pythoninline{dH} and \pythoninline{drho} are the trial functions for the enthalpy and density functions and are not utilized in this code, but are included for reference later on.

We define the complete non-linear residual as the sum of equations (13), (14), and (15): 
\begin{align*}
    f = f_H + f_{\rho} + f_w.
\end{align*}
Solving this system can be accomplished with \emph{Newton's Method} which requires derivation of the Jacobian:
\begin{align*}
    J = \frac{\partial f}{\partial d_h}.
\end{align*}
In FEniCS this is done with:
\begin{python}
f  = f_H + f_rho + f_w
J  = derivative(f, h, dh)
\end{python}

\section{Age Solution}

The age of firn is described with the equation
\begin{align}
  \frac{\partial a}{\partial t} = 1 - w \nabla{a}.
\end{align}
Because this equation is purely advective, upwinding is needed; the method chosen here is \emph{Taylor-Galerkin} and is described in \emph{Codina} [1997].  The Taylor series expansion of $w$ is
\begin{align*}
  a^{k+1} = a^{k} + \frac{\partial a^k}{\partial t}\Delta t + 
            \frac{1}{2}\frac{\partial^2 a^{k+\theta}}{\partial t^2}\Delta t^2 + 
            \mathrm{O} (\Delta t^3).
\end{align*}
Replacing the time partials with the previous equation and eliminating the $\Delta t^3$ term, we have
\begin{align*}
  a^{k+1} = a^{k} + \left[1 - w \nabla{a}^k \right] \Delta t
            - \frac{1}{2}\frac{\partial}{\partial t}
              \left[1 - w \nabla{a}^{k+\theta} \right] \Delta t^2. 
\end{align*}
Rearranging and simplifying,
\begin{align*}
  \frac{a^{k+1} - a^{k}}{\Delta t} = 1 - w \nabla{a}^k
        - \frac{w \Delta t}{2} 
          \left( \nabla \frac{\partial a^{k+\theta}}{\partial t} \right).
\end{align*}
The time differential in the last term can again be replaced, 
\begin{align*}
  \frac{a^{k+1} - a^{k}}{\Delta t} = 1 - w \nabla{a}^k
        - \frac{w \Delta t}{2} 
          \left( \nabla [1 - w \nabla a^{k+\theta}] \right),
\end{align*}
reduced to
\begin{align*}
  \frac{a^{k+1} - a^{k}}{\Delta t} &= 1 - w \nabla{a}^k
        + \frac{w \Delta t}{2} 
          \left( \nabla [w \nabla a^{k+\theta}] \right),
\end{align*}
and finally evaluated using the product rule as
\begin{align*}
  \frac{a^{k+1} - a^{k}}{\Delta t} = &1 - w \nabla{a}^k\\
        &+ \frac{w \Delta t}{2} 
          \left( w \nabla^2 a^{k+\theta} + \nabla w \nabla a^{k+\theta} \right).
\end{align*}
After moving all the terms to one side, multiplying by the test function $\xi$, and integrating over $\Omega$ we have
\begin{align*}
  f_a = \int_{\Omega} \frac{a^{k+1} - a^{k}}{\Delta t}\xi\ d\Omega - 
      \int_{\Omega} 1\xi\ d\Omega
  + \int_{\Omega} w \nabla{a}^k \xi\ d\Omega \\
  - \int_{\Omega} \frac{w^2 \Delta t}{2} 
    \nabla^2 a^{k+\theta} \xi\ d\Omega.
  - \int_{\Omega} \frac{w \Delta t}{2} 
    \nabla w \nabla a^{k+\theta} \xi\ d\Omega.
\end{align*}
Integrating the second to last term by parts and disregarding the boundary term results in
\begin{align*}
  f_a = \int_{\Omega} \frac{a^{k+1} - a^{k}}{\Delta t}\xi\ d\Omega - 
      \int_{\Omega} 1\xi\ d\Omega
  + \int_{\Omega} w \nabla{a}^k \xi\ d\Omega \\
  + \int_{\Omega} \frac{w^2 \Delta t}{2} 
    \nabla a^{k+\theta} \cdot \nabla \xi\ d\Omega.
  - \int_{\Omega} \frac{w \Delta t}{2} 
    \nabla w \nabla a^{k+\theta} \xi\ d\Omega.
\end{align*}
This final residual may be created in FEniCS as:

\begin{python}
a     = Function(V)
da    = TrialFunction(V)
xi    = TestFunction(V)
a_1   = Function(V)

theta = 0.5
a_mid = theta*a + (1-theta)*a_1
f_a   = (a - a_1)/dt*xi*dx 
        - 1.*xi*dx 
        + w*grad(a_mid)*xi*dx 
        + w**2*dt/2*grad(a_mid)*grad(xi)*dx 
        - w*grad(w)*dt/2*grad(a_mid)*xi*dx
\end{python}

\section{Boundary Conditions}

A cyclical enthalpy boundary condition for the surface can be simulated with 
\begin{align*}
    H_s &= c_i ( T_s - T_0 ),\\
    T_s &= T_{avg} + \alpha \sin(\gamma t),
\end{align*}
where $\alpha$ is the amplitude of temperature variation and $\gamma = 2\pi / spy$ is the frequency.  The surface-density boundary condition can be likewise described as (see Figure 1): 
\begin{align*}
    \rho_s^{n+1} = \rho_{\dot{b}}^{n+1} d_p + \rho_s^{n} (1 - d_p).
\end{align*}
The velocity of the surface can be described as 
\begin{align*}
  w_s = -\frac{\dot{b}}{\rho_s},\ \dot{b} = \rho_i \dot{a},
\end{align*}
Where $\dot{a}$ is the accumulation rate in m a\sups{-1}.
The age of the firn on the surface will always be 0.
Each of these of these can be created with FEniCS by
\begin{python}
code = 'c*(Tavg + 9.9*sin(omega*t) - T0)'
Hs   = Expression(code, c=cp, Tavg=Tavg, 
                  omega=freq, t=t0, T0=T0)

code = 'dp*rhon + (1 - dp)*rhoi'
rhoS = Expression(code, rhon=rhosi, 
                  rhoi=rhosi, dp=1.0)

code   = '- (rhoi * adot / spy) / rhos'
wS     = Expression(code, rhoi=rhoi, 
                    adot=adot, spy=spy, 
                    rhos=rhos)

ageS   = Constant(0.0)

def surface(x, on_boundary):
  return on_boundary and x[0] == zs

Hbc   = DirichletBC(MV.sub(0), Hs,   surface)
Dbc   = DirichletBC(MV.sub(1), rhoS, surface)
wbc   = DirichletBC(MV.sub(2), wS,   surface)
ageBc = DirichletBC(V,         ageS, surface)
\end{python}
Within the time-loop the variables in each \pythoninline{Expression} object can be updated as needed by utilizing the dot operator.

Now all that is left is to iterate through time and call the \pythoninline{solve} method at each step:\par
\begin{python}
solve(f == 0, h, [Hbc, Dbc, wbc], J=J)
solve(f_a == 0, a, ageBc)
\end{python}
The \pythoninline{solve} function chooses \emph{Newton's Method} by default to solve the non-linear enthalpy, density, and velocity residual by minimizing \pythoninline{f}.  The boundary conditions are updated with these calls by specifying the list \pythoninline{[Hbc, Dbc, wbc]} and \pythoninline{ageBc} in solver parameters.

\begin{figure}[H]
  \centering
    \includegraphics[width=0.50\textwidth]{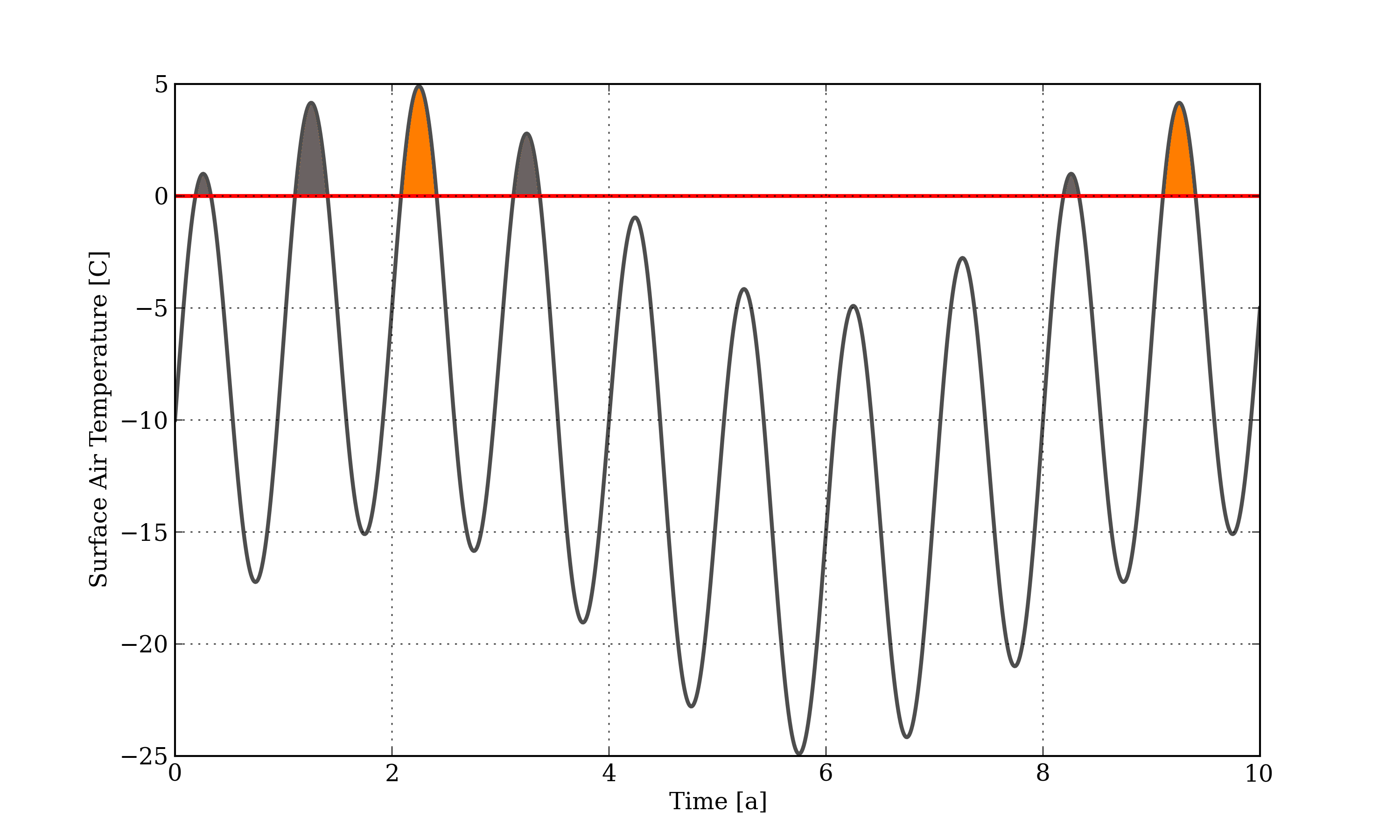}
	\caption{\footnotesize The surface 2-meter air temperature using the equation above with $\alpha = 10$ and $\beta = 5$.  Temperatures above 0$^{\circ}$ C are shaded.}
\end{figure}

\section{Model Parameters}

Within the time-loop there are a number of parameters which need to be updated.  Taking into account conservation of mass, the height $l$ of each node must be re-calculated:\par
\begin{align*}
  l_{new} = l_{ini} \frac{\rho_{ini}}{\rho},
\end{align*}
where $\rho_{ini}$ abd $l_{ini}$ are the density and height vectors of the firn column when the system was initialized.  With the height of the nodes calculated, the z-positions may be found by iterating through the heights and setting the z vector's corresponding cell equal to the current sum.  These tasks may be completed with the following code performed by the \pythoninline{update\_height} function of the \pythoninline{firn} class:\par
\begin{python}
avg_rhoin = (self.rhoin[1:] + self.rhoin[:-1])
            / 2
avg_rho   = (self.rho[1:] + self.rho[:-1]) / 2
lnew      = self.lini * avg_rhoin / avg_rho
zSum      = self.zb
zTemp     = np.zeros(self.n)
for i in range(self.n)[1:]:
  zTemp[i] = zSum + lnew[i-1]
  zSum    += lnew[i-1]
self.z    = zTemp
self.l    = lnew
index     = self.index
self.mesh.coordinates()[:,0][index] = self.z
\end{python}
The variable \pythoninline{index} is an array of positions corresponding to the correct ordering of the nodes, necessary after mesh refinement.  

The height $z_o$ at time index $n$ of the original surface may be calculated as follows:
\begin{align*}
  z_o^{n} = (z_s^0 - z_b) \frac{z_o^{n-1} - z_b}{z_s^{n-1} - z_b}
           + w_{z_o} \Delta t.
\end{align*}
This maintains the relative location of the original surface to the current surface and moves downward proportional to $w$.  This is accomplished in the class named \pythoninline{firn} with\par
\begin{python}
interp  = interp1d(self.z, self.w,
                   bounds_error=False,
                   fill_value=self.w[0])
zint    = np.array([self.zo])
wzo     = interp(zint)[0]
dt      = self.dt
zs      = self.z[-1]
zb      = self.z[0]
zs_1    = self.zs_1
zo      = self.zo
self.zo = (zs - zb) * (zo - zb) / (zs_1 - zb) 
          + wzo * dt
\end{python}
The indexes to \pythoninline{self.z} refers to the surface, \pythoninline{[-1]}, or the base, \pythoninline{[0]}.  For all operations it is convenient to store all the state data from the simulation in an object for ease of access.  It was for this purpose the \pythoninline{firn} class was created and contains the signature\par 
\begin{python}
firn(data, z, l, index, dt)
\end{python}
with \pythoninline{data} a tuple containing the main variables, \pythoninline{z} the node z-positions as defined above, \pythoninline{l} the height vector of the elements, \pythoninline{index} the index of re-ordered mesh locations, and \pythoninline{dt} the time step of the program, $\Delta t$.

The variables for accumulation and surface temperature are the main driving forces in the simulation, and data from a specific site may be used in the model by interpolating the data in increments of $\Delta t$ and inserting the values into the equations.  This may be accomplished with the \pythoninline{set\_local(n)} method of the \pythoninline{vector} class, which takes as input a NumPy array \pythoninline{n} with indexes corresponding to node positions within the \pythoninline{mesh} object.  If the variable is used in the surface boundary condition, this may be updated within the FEniCS \pythoninline{Expression} object with the dot operator.

A function has been provided (\pythoninline{set\_ini\_conv}) which initializes the density to a previously derived density.  The density may also be initialized to a set of real-world data if desired, and is demonstrated in the temperature equation model, \texttt{objModel.py}.

The \texttt{plot.py} file contains the class \pythoninline{firn} and the previously undescribed \pythoninline{plot} class.  This class uses the plotting package MatPlotLib to display the data contained in the \pythoninline{firn} object.  The method \pythoninline{plot\_all\_height()} plots the height history for a group of simulations and is useful for comparing the effects of model parameters. 

Another version of \pythoninline{Model}, the main simulation class, has been created which uses collected data for density and surface temperature.  When using this version, it is important to make $\Delta t$ less than or equal to the time interval of recorded events so all data points are included.

\section{Variable Definitions}

Many variable are used in this simulation and many do not change.  These are defined below:\\

\noindent\textbf{Constants :}
\begin{center}
\footnotesize
\noindent\begin{tabular}{lccc}
\hline
Var. & Value & Units & Description\\
\hline
$g$ & $9.81$ & m s\sups{-2} & gravitational acceleration\\
$R$ & $8.3144621$ & J mol\sups{-1} K\sups{-1} & gas constant\\
$spy$  & $31556926$ & s & seconds per year\\
$\rho_i$ & $917$ & kg m\sups{-3} & density of ice\\
$\rho_w$ & $1000$ & kg m\sups{-3} & density of water\\
$\rho_m$ & $550$ & kg m\sups{-3} & critical density value\\
$k_i$  & $2.1$ & W m\sups{-1}K\sups{-1} & thermal conductivity of ice\\
$c_i$  & $2009$ & J kg\sups{-1}K\sups{-1} & heat capacity of ice\\
$L_f$ & \SI{3.34e5} & J kg\sups{-1} & latent heat of fusion\\
$H_s$ & $c_i(T_w - T_0)$ & J kg\sups{-1} &  Enthalpy of ice at $T_w$\\
$T_w$  & $273.15$ & K & triple point of water\\
$T_0$ & $0.0$ & K & reference temperature\\
$k_g$ & \SI{1.3e-7} & m\sups{2}s\sups{-1} & grain growth coefficient\\
$E_c$ & \SI{60e3} & J mol\sups{-1} & act. energy for water in ice\\
$E_g$ & \SI{42.4e3} & J mol\sups{-1} & act. energy for grain growth\\
\hline
\end{tabular}
\normalsize\\
\end{center}

\noindent Variables used by the model can be specified to suit simulation requirements:\\

\noindent\textbf{Model Specific :}
\begin{center}
\footnotesize
\noindent\begin{tabular}{lcc}
\hline
Var. & Units & Description\\
\hline
$\rho_s$ & kg m\sups{-3} & initial density at surface\\
$\dot{a}$ & m a\sups{-1} & ice eq. surface accumulation rate\\
$\dot{b}$  & kg m\sups{-2}s\sups{-1} & surface accumulation\\
$A$  & mm a\sups{-1} & surface accumulation\\
$V_a$  & m s\sups{-1} & mean annual wind speed\\
$T_{avg}$ & K & average annual temperature\\
$T_{s}$ & K & firn surface temperature\\
$z_s$ & m & surface start z-location\\
$z_b$ & m & firn base z-location\\
$z_{s_1}$ & m & previous time-step's surface\\
$dz$ & m & initial z-spacing\\
$l$ & m & vector of node heights\\
$\Delta t$ & s & time-step\\
$t_0$ & s & begin time\\
$t_f$ & s & end-time\\
\hline
\end{tabular}
\normalsize\\
\end{center}

\section{Verification of Program}

A converging run of the program was done quickly by making $\Delta t$ equal $spy$: this has the effect of producing a steady-state solution.  After the density-profile converged the data was saved to a text file in the \pythoninline{data} folder.  The script was run again with the average surface air temperature $T_{avg}$ made so   that the surface temperature peaks at $8^{\circ}$ C if $t < 10$ years, and $0^{\circ}$ C if $t \geq 10$ years. This has the effect of halting any melting and refreezing after this time period.  For this run the method \pythoninline{set\_ini\_conv} was called to initialize the previous runs data and $\Delta t$ was set to $0.0025*spy$; the results are shown in Figures 3 and 4.

\section{Interpretation}

The surface-density equation has two values for new accumulation: $\rho_s$ and $\rho_w$ depending on the 2-meter average surface air temperature.  This is quite simplified; a better approach that models real-life circumstances can be found.  

Testing with real-world temperature and accumulation data-sets is required to validate the model.  The simulation's surface-height- and density-profile outputs while using these data may then be compared against cataloged surface-height and density-profile data to verify its accuracy.

Above the ice lens in Figure 3 you will see numerical distortion: this is caused by the sudden rise in density where the lens begins.  This distortion is a source of inaccuracies and needs also to be corrected.

\section{Work in Progress}
At its current state of development the model does not take into account water transport through the firn column, describable with the Darcy flow equations:
\begin{align*}
    q = \frac{-k}{\mu}\nabla P, \ \  v = \frac{q}{\phi},
\end{align*}
where $k$ is permeability, $\mu =$ \SI{1.787e-3} Pa$\cdot$s is the viscosity of water at $0^{\circ}$ C, $\nabla P$ is the pressure gradient vector, and $\phi = 1- \rho/\rho_i$ is porosity.

\emph{Waldner et al.} [2002] introduces this issue and provides references to numerous models which simulate this phenomenon.  \emph{Coleou et al.} [1998] supplies an equation for the irreducible water content of snow
\begin{align*}
    S_0 = \frac{0.0057}{1 - \phi} + 0.0017,
\end{align*}
which may be used with the expression for permeability from \emph{Bozhinskiy \& Krass} [1989] :
\begin{align*}
    k = k_0 \exp(m \phi)\left( \frac{S - S_0}{\phi - S_0} \right)^2,
\end{align*}
where $S$ is the relative water content and $k$ \& $m$ are empirical constants.

\section{Concluding Remarks}
This model is a good start towards accurately modeling the densification of firn: it uses the work of many established models and has the potential to expand;  The simulation is able to assimilate data easily, is mathematically easy to interpret, and runs efficiently.  This subject is a worthy candidate for further study.  

\begin{figure}[H]
	\centering
		\includegraphics[width=0.42\textwidth]{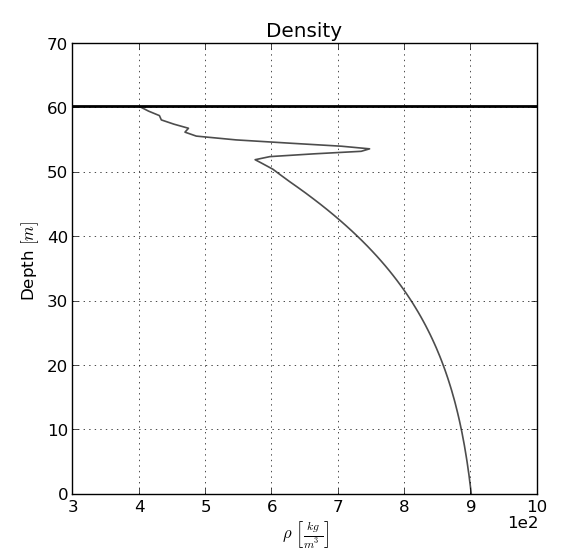}
	\label{fig:500 year orbit}
	\caption{\footnotesize Density profile after 40 years with an ice lens approximately 7 meters below the surface}
\end{figure}

\begin{figure}[H]
	\centering
		\includegraphics[width=0.42\textwidth]{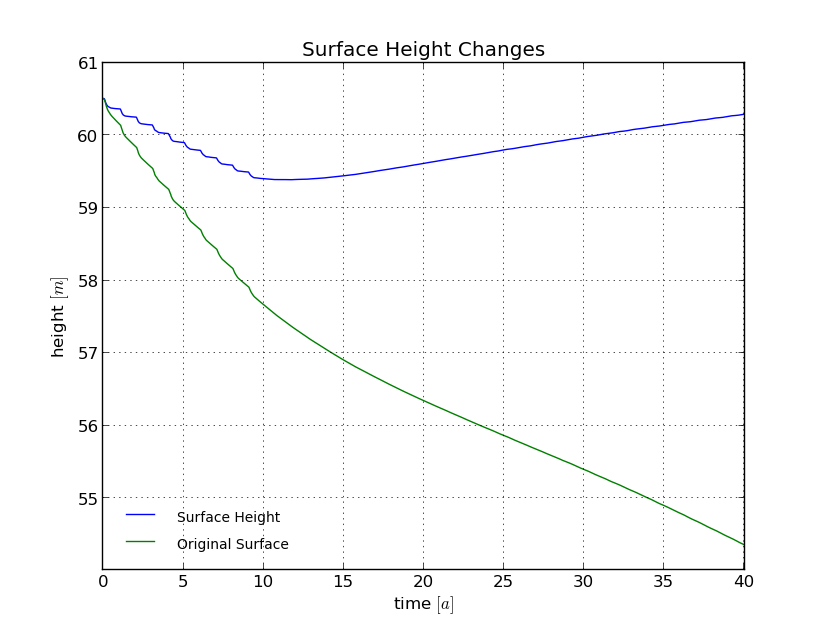}
	\label{fig:500 year orbit}
	\caption{\footnotesize 40-year height history of the column (blue) and original surface (green) resulting in the previous figure.  This shows a rapid decrease in height as the lens is formed in the first ten years of simulation.  The fluctuations in height show an increase in height with winter temperatures.}
\end{figure}

\begin{figure}[H]
  \centering
    \includegraphics[width=0.50\textwidth]{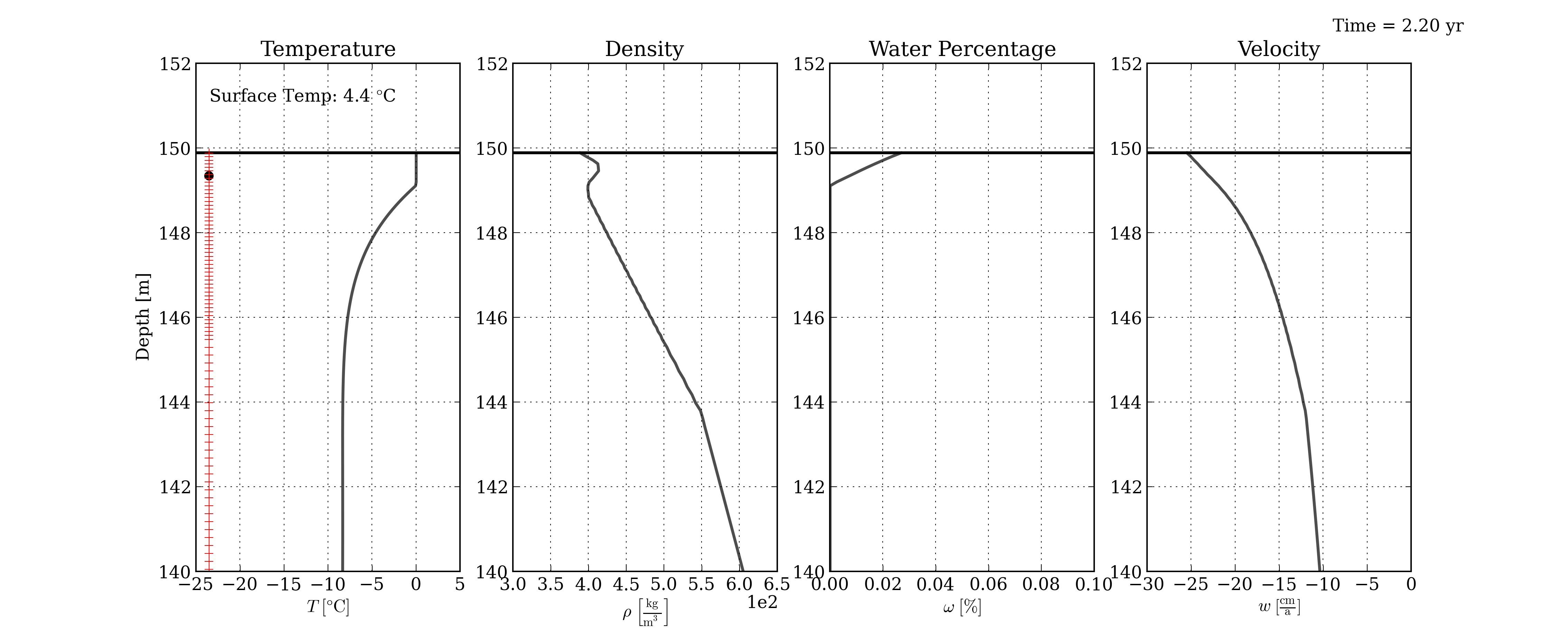}
	\caption{\footnotesize Simulation output at $t = 2.2$ years corresponding to the first orange spike in Figure 1.  The red `+' symbols show element intersections, and the black dot follows the original surface height when the simulation started.  As time moves forward the original surface is covered by new layers of snow and moves down into the column.  Also notice the change in densification rate at 550 kg m\sups{-3}.}
\end{figure}

\begin{figure}[H]
  \centering
    \includegraphics[width=0.50\textwidth]{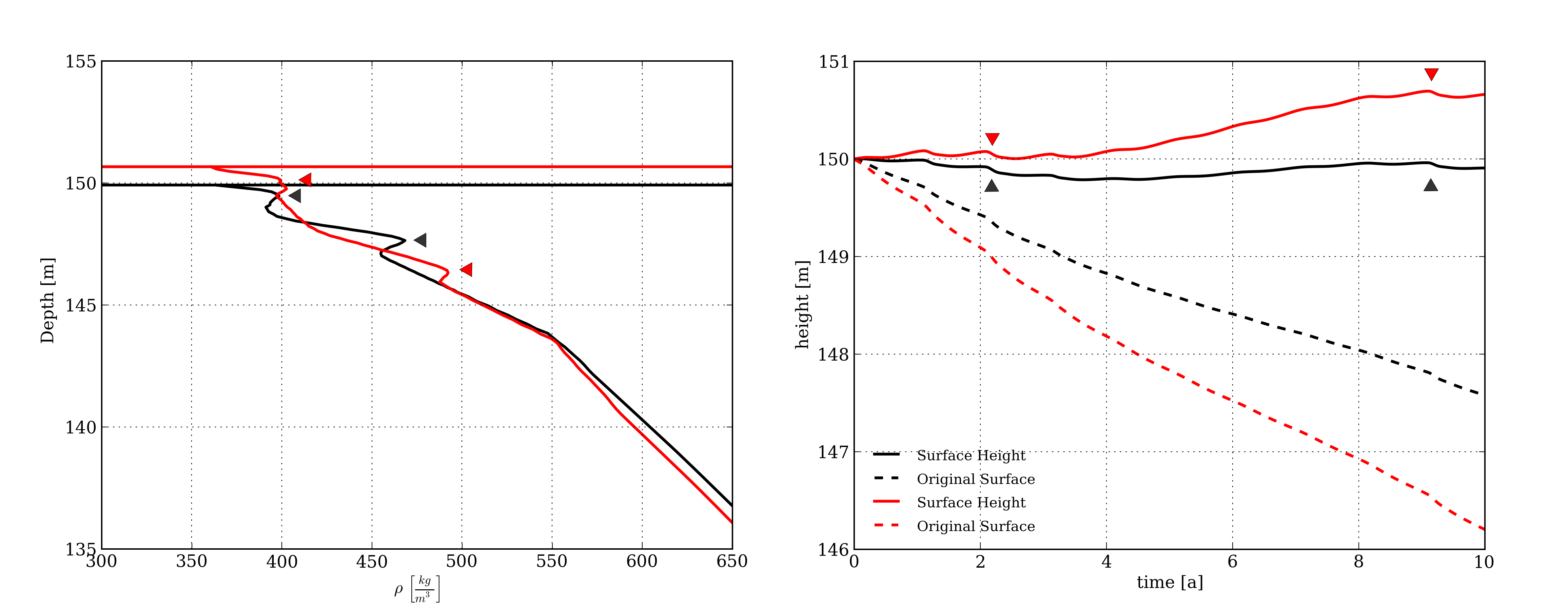}
	\caption{\footnotesize Density profiles (left) and height history (right) after 10 years.  The black line has $\dot{a} = 0.10$ m a\sups{-1} and the red line has $\dot{a} = 0.20$ m a\sups{-1}.  Triangles show the density increases and height decreases due to the orange peaks in Figure 2.}
\end{figure}

\newpage

\end{document}